\documentclass[sigplan,twocolumn,nonacm]{acmart}

\renewcommand\footnotetextcopyrightpermission[1]{}
\usepackage{booktabs}
\usepackage{tikz}
\usepackage{comment}
\usepackage{amsmath}
\usepackage{caption}
\usepackage{subcaption}
\usepackage{graphicx}
\usepackage{float}
\usepackage{algorithm}
\usepackage{algorithmicx}
\usepackage[noend]{algpseudocode}
\algnewcommand{\LineComment}[1]{\State \(//\) #1}
\usepackage[normalem]{ulem}
\usepackage{bbding}
\usepackage{todonotes}
\usepackage{filecontents}
\usepackage{enumitem}
\usepackage[bottom]{footmisc}
\usepackage[]{hyperref}
\usepackage{cleveref}
\usepackage{pifont}
\usepackage{multirow}
\usepackage{xcolor}
\usepackage{outlines}
\usepackage{diagbox}
\usepackage{wasysym}

\definecolor{LightGray}{gray}{0.9}
\definecolor{FadedBanana}{RGB}{255,255,191}
\definecolor{DeepChalk}{RGB}{255,191,191}
\definecolor{FadedFlora}{RGB}{191,255,191}
\definecolor{DeepSnow}{RGB}{191,255,255}
\definecolor{SoapStone}{RGB}{218,218,218}
\definecolor{LightCayenneSixty}{RGB}{239,206,211}
\definecolor{LightCayenne}{RGB}{208,143,145}

\newcommand{\sysname}{{H\textsc{arli}}}

\begin{document}

\title[]{\sysname{}: SLO-Aware Co-location of LLM Inference and PEFT-based Finetuning on Model-as-a-Service Platforms}

\author{Ao Xu}
\authornote{Both authors contributed equally to this research.}
\affiliation{
  \institution{Shanghai Jiao Tong University}
  \country{}
}
\author{Han Zhao}
\authornotemark[1]
\affiliation{
  \institution{Shanghai Jiao Tong University}
  \country{}
}
\author{Weihao Cui}
\affiliation{
  \institution{Shanghai Jiao Tong University}
  \country{}
}
\author{Quan Chen}
\authornote{Quan Chen is the corresponding author.}
\affiliation{
  \institution{Shanghai Jiao Tong University}
  \country{}
}
\author{Yukang Chen}
\affiliation{
  \institution{Shanghai Jiao Tong University}
  \country{}
}
\author{Shulai Zhang}
\affiliation{
  \institution{Shanghai Jiao Tong University}
  \country{}
}
\author{Shuang Chen}
\affiliation{
  \institution{Shanghai Jiao Tong University}
  \country{}
}
\author{Jiemin Jiang}
\affiliation{
  \institution{Zhejiang University}
  \country{}
}
\author{Zhibin Yu}
\affiliation{
  \institution{Chinese Academy of Science}
  \country{}
}
\author{Minyi Guo}
\affiliation{
  \institution{Shanghai Jiao Tong University}
  \country{}
}

\begin{abstract}
Large language models (LLMs) are increasingly deployed under the Model-as-a-Service (MaaS) paradigm. To meet stringent quality-of-service (QoS) requirements, existing LLM serving systems disaggregate the prefill and decode phases of inference. However, decode instances often experience low GPU utilization due to their memory-bound nature and insufficient batching in dynamic workloads, leaving compute resources underutilized. 

We introduce \sysname{}, a serving system that improves GPU utilization by co-locating parameter-efficient finetuning (PEFT) tasks with LLM decode instances. PEFT tasks are compute-bound and memory-efficient, making them ideal candidates for safe co-location. Specifically, \sysname{} addresses key challenges—limited memory and unpredictable interference—using three components: a unified memory allocator for runtime memory reuse, a two-stage latency predictor for decode latency modeling, and a QoS-guaranteed throughput-maximizing scheduler for throughput maximization. Experimental results show that \sysname{} improves the finetune throughput by 46.2\% on average (up to 92.0\%) over state-of-the-art serving systems, while maintaining strict QoS guarantees for inference decode.

\end{abstract}

\maketitle

\section{Introduction}
\label{sec:intro}



Large language models (LLMs) \cite{openai2023gpt4, deepseekai2025deepseekr1incentivizingreasoningcapability, touvron2023llamaopenefficientfoundation, meta-llama2023llama3, qwen, genimi} have greatly improved application performance across a wide range of domains. As a result, LLM inference has become one of the most critical and widely deployed user-facing services, commonly provided under the Model-as-a-Service (MaaS) paradigm.

To meet service quality requirements in MaaS, existing LLM inference systems typically adopt a disaggregated deployment strategy.
This is because the phases for processing LLM requests—namely, the prefill and decode phases—have distinct quality-of-service (QoS) targets\footnote{The QoS target refers to time-to-first-token (TTFT) for the prefill phase and time-per-output-token (TPOT) for the decode phase.} and compute characteristics \cite{distserve,splitwise}.
Without disaggregation, the two phases interleave and may cause QoS violations for each other.
Disaggregating inference instances into prefill and decode instances eliminates this interleaving.
\begin{figure}
\centering
\includegraphics[width=\linewidth]{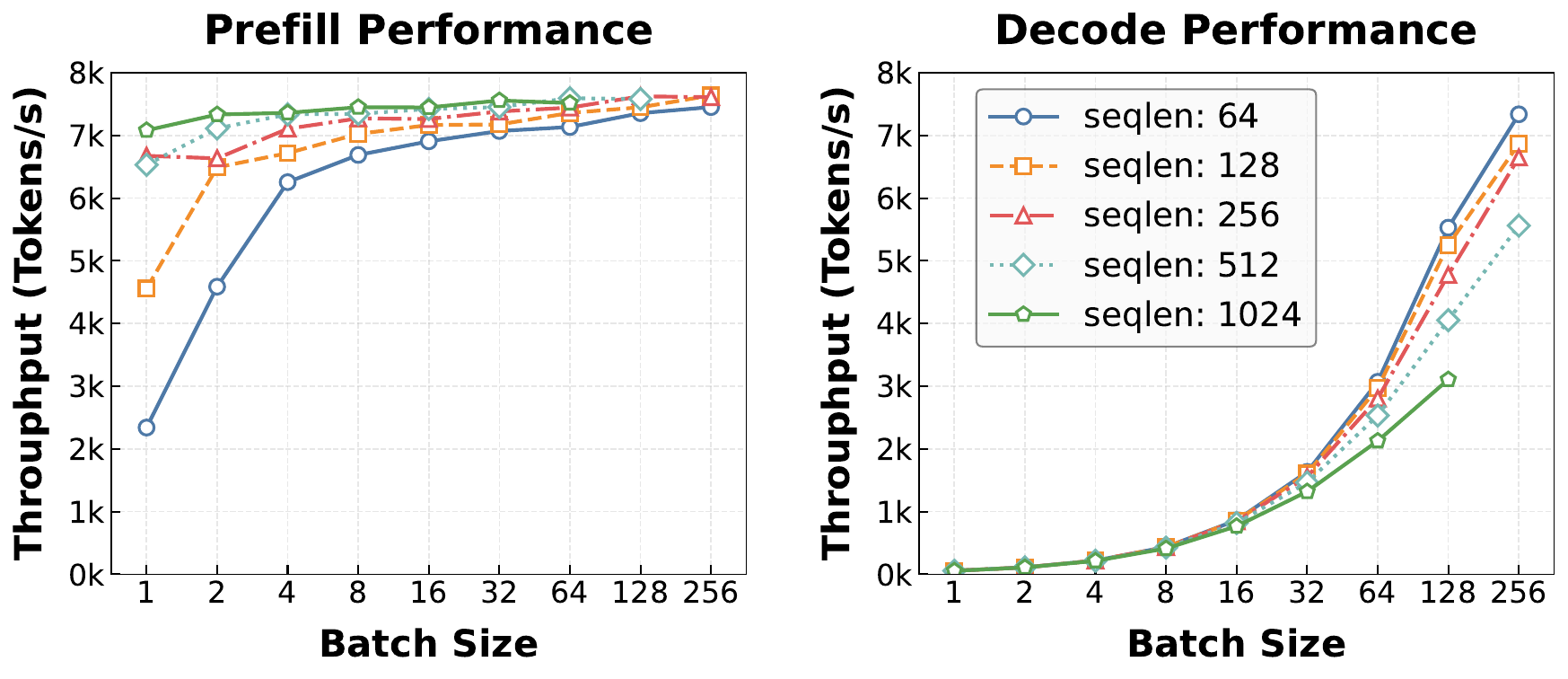}
\vspace{-4mm}
\caption{Throughput of prefill and decode phase with varied batch sizes and sequence lengths.}
\vspace{-4mm}
\label{fig:pd_bs}
\end{figure}

However, we find that decode instances can easily suffer from low utilization in disaggregated deployments for online serving.
\autoref{fig:pd_bs} shows the throughput of the prefill and decode phases across different batch sizes ($bs$) and sequence lengths ($seqlen$).
The throughput of the prefill phase quickly flattens as $bs$ increases.
When $seqlen$ is large (e.g., 1024), the prefill phase may begin to flatten even at $bs = 1$.
In contrast, the decode phase does not reach saturation even when $bs$ approaches 256.
This indicates that while LLM request can easily saturate the prefill instance, decode instances depend heavily on larger batch sizes to increase the execution parallelism for high hardware utilization.
Given the highly dynamic runtime request patterns in LLM inference workloads~\cite{splitwise}, decode instances often lack sufficiently large batch sizes, leading to wasted GPU resources.

Our deeper investigation in \S\ref{sec:problem} shows that decode phases are memory-bound at small $bs$, leading to underutilized compute resources.
To harvest these wasted resources in decode instances, a natural approach is to allocate a subset of SMs (streaming multiprocessors—the compute units of a GPU) spatially to a compute-intensive task, thereby maximizing both compute and bandwidth utilization.
However, such a co-location approach faces significant challenges in LLM serving systems due to limited memory capacity and unpredictable execution interference.

Firstly, a single GPU may not have sufficient memory to co-locate another task for harvesting idle resources.
For example, compute-intensive workloads like traditional LLM training can fully occupy GPU memory.
Secondly, unpredictable interference can lead to QoS violations.
Since the decode workload varies significantly during online serving, co-locating another dynamic task like the decode workload makes interference difficult to predict.
To safely harvest unused GPU resources from decode instances during online LLM inference, it is essential to carefully select a suitable co-located workload.

After a thorough investigation, we identify PEFT-based (parameter-efficient finetuning) LLM finetuning tasks as promising candidates for co-location with LLM decode instances.
Notably, PEFT services are also pervasive workloads, as leading MaaS providers such as OpenAI offer them to support user-customized LLM adaptations\cite{openai-fine-tuning}.
PEFT is a good fit in two key aspects.
1) PEFT is compute-bound but updates only a small fraction of model weights. Compared to pretraining or full-parameter finetuning, it consumes significantly less GPU memory than traditional LLM training tasks.
2) LLM finetuning tasks tend to be relatively stable, typically using fixed batch sizes. As a result, their performance interference on LLM inference services can be modeled and quantified through linear regression.

To this end, we propose \sysname{}, an LLM serving system that efficiently co-locates inference and finetune tasks on the same GPUs.
\sysname{} successfully harvest the wasted GPU resources in decode instances for online serving with QoS guarantees.
To achieves this, it comprises three components: a {\it unified memory allocator}, a {\it two-stage latency predictor}, and a {\it QoS-guaranteed throughput-maximizing scheduler}. 

While LLM finetuning is already memory-efficient and compute-intensive, it remains infeasible to directly co-locate it with decode phases.
This is because existing serving systems reserve a memory pool to store the intermediate KV cache required for generating new output tokens.
The unified memory allocator enables inter-task memory sharing.
It repurposes unused slots in this memory pool to satisfy the memory demands of LLM finetuning tasks via low-level memory mapping~\cite{nvidia-vmm}.
It also integrates a window-based memory swapping mechanism to further mitigate the memory pressure that finetuning imposes on the decode phase.

\sysname{} spatially co-locates finetuning and decode instances by partitioning SMs using low-level GPU resource management mechanisms~\cite{greencontext}.
Leveraging the stability of PEFT, the latency predictor estimates the current decode latency and projects performance under different SM partitioning schemes in a two-stage process.
Based on these predictions, the scheduler dynamically adjusts SM allocations to ensure inference QoS while maximizing finetuning throughput under varying load conditions.






We implement \sysname{} on top of SGLang~\cite{sglang}, extending it with our proposed co-location mechanisms. \sysname{} is evaluated on mainstream LLMs (LLaMA3-8B and Qwen2.5-7B) using a production LLM service trace\cite{splitwise}. Experimental results show that \sysname{} improves finetuning throughput by 46.2\% on average (up to 92.0\%) over state-of-the-art baselines, while consistently meeting inference QoS requirements. We plan to open-source \sysname{} after publication.



In summary, we make the following contributions:
\begin{itemize}
[leftmargin=*]
    \item We identify LLM finetune tasks as a suitable candidate for co-location with LLM inference services to improve system throughput and GPU resource utilization.
    \item We propose a unified memory allocator that enables finetune tasks to leverage unused memory from inference services, supporting efficient execution under tight memory constraints.
    \item We develop a linear-regression-based latency predictor that models inference performance under bandwidth contention, enabling adaptive resource allocation for co-located tasks under dynamic workloads.
\end{itemize}
\section{Background and Motivation}


\begin{figure}
\centering
\includegraphics[width=\linewidth]{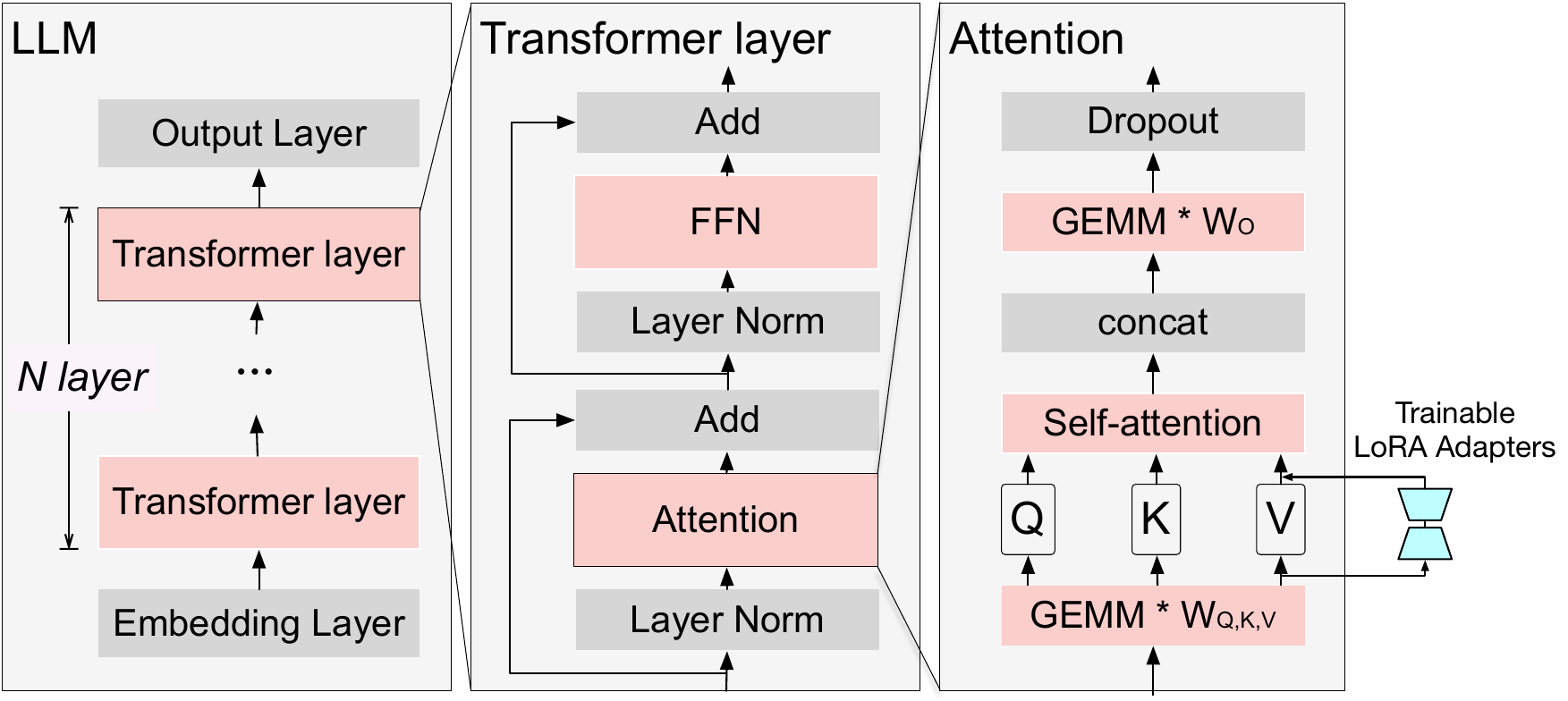}
\caption{The general architecture of an LLM model.}
\label{fig:llm}
\end{figure}

\subsection{LLM inference and LLM finetune}
\autoref{fig:llm} illustrates the general architecture of an LLM, which consists of an embedding layer, a stack of transformer layers, and an output layer.
Each transformer layer consists of two primary components: the self-attention module and the feed-forward network (FFN). 
Deep inside, these operators mainly perform matrix multiplications with varying shapes.

As for online serving, the inference process for an LLM request typically consists of two distinct phases: the prefill phase, which computes the first token based on the provided prompt, and the decode phase, which performs autoregressive generation of subsequent tokens.
The two phases are consecutive: the intermediate results generated during the prefill phase, known as the KV cache, are shared with the subsequent decode phases to avoid redundant computation.
As mentioned in \S\ref{sec:intro}, state-of-the-art serving systems~\cite{NEURIPS2024_724be447,kwon_efficient_2023} often disaggregate the two phases into different instances and processing them with independent batch size to meet their distinct QoS target requirements. 


LLM finetuning involves updating model parameters based on training data. Each iteration consists of a forward pass, a backward pass, and gradient update.
Unlike inference, finetuning requires storing intermediate activations for gradient computation, leading to higher memory usage.
In full-parameter finetuning, all model parameters are trainable, imposing significant computational and memory pressure when adapting a pretrained LLM for specific tasks.

To address this, researchers have proposed parameter-efficient finetuning (PEFT) techniques, which freeze most parameters and train only a small subset.
As shown on the right of \autoref{fig:llm}, LoRA is a well-known PEFT technique that freezes all parameters of the original model and trains only the attached LoRA adapters (less than $0.3\%$ of the total parameters).
As a result, PEFT significantly reduces memory usage compared to full-parameter finetuning.
Notably, leading MaaS providers, such as OpenAI\cite{openai-fine-tuning} and Hugging Face\cite{huggingface-peft} support model customization through PEFT.

\subsection{Problems and Opportunities}\label{sec:problem}

\subsubsection{Characterizing Decode Instances}

\begin{figure}
\centering
\includegraphics[width=\linewidth]{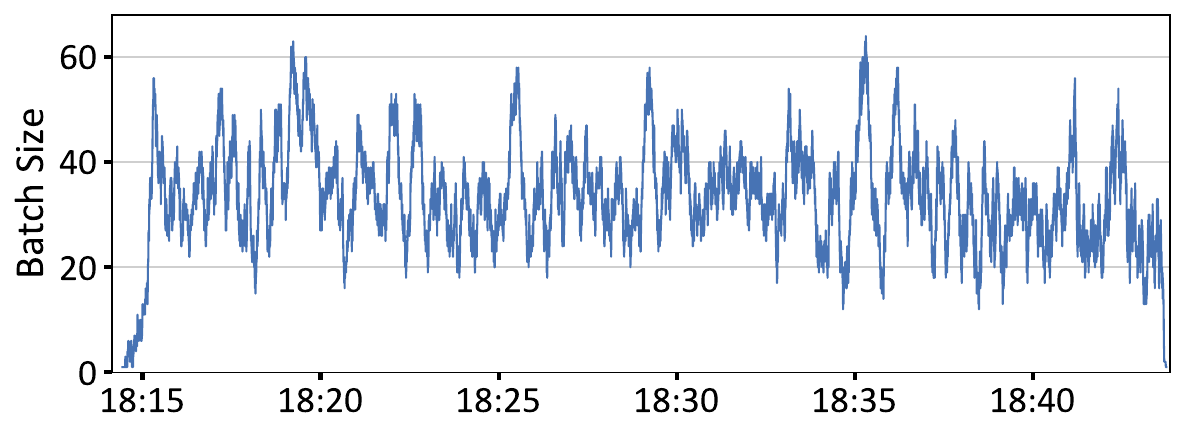}
\caption{The decode batch size of inference tasks under a real-world trace~\cite{splitwise}.}
\label{fig:trace bs}
\end{figure}

As mentioned in \autoref{sec:intro}, the decode phase is particularly reliant on batch processing to improve system throughput and GPU utilization.
We further profile the runtime batch size of a mainstream LLM serving system, SGLang\cite{sglang}, under real-world workload traces\cite{splitwise}. Specifically, we use an open-source LLM request trace released by Microsoft. \autoref{fig:trace bs} shows the runtime batch size of SGLang running on an Nvidia Ada6000 GPU. 
The batch size demonstrates significant variability. This is due to runtime load fluctuations, which are inherent in user-facing services. As a result, LLM inference frequently operates under small batch sizes, leading to low system throughput and suboptimal GPU utilization.

Next, we analyze the DRAM bandwidth utilization and SM utilization during the decode phase across different configurations. Specifically, we identify the major GPU kernels involved in the decode phase and recorde their execution times. We then use $ncu$ to collect kernel-level DRAM and SM utilization metrics and compute the overall hardware utilization of the decode phase according to \autoref{eq:Util}. In this equation, $Util_{SM-k_i}$ denotes the SM utilization of a kernel, $Util_{DRAM-k_i}$ denotes the bandwidth utilization of the kernel, and $R_{k_i}$ represents the proportion of the kernel’s execution time relative to the total decode phase duration.
\begin{equation}
\small
\begin{aligned}
    Util_{SM-decode} &= \sum Util_{SM-k_i} \times R_{k_i} \\
    Util_{DRAM-decode} &= \sum Util_{DRAM-k_i} \times R_{k_i} \\
    R_{k_i} &= T_{k_i} / T_{overall}
\end{aligned}
\label{eq:Util}
\end{equation}

\begin{figure}
\centering
\includegraphics[width=\linewidth]{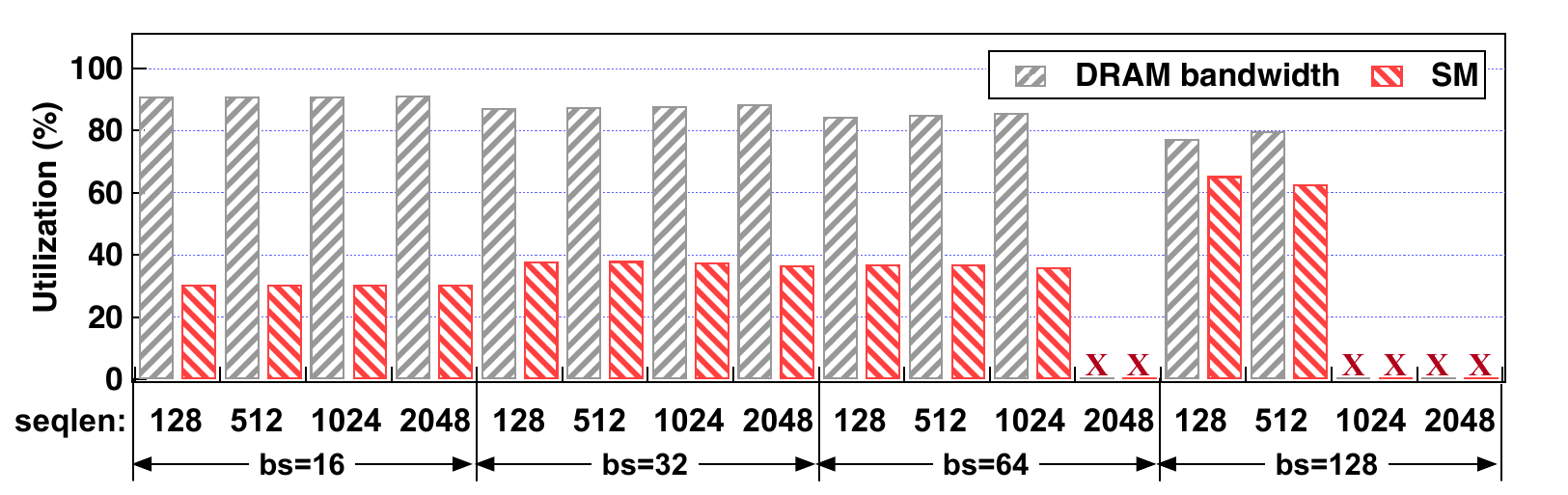}
\caption{The DRAM bandwidth and SM utilization of the decode phase under different configurations.}
\label{fig:fexp-moti2}
\end{figure}

\autoref{fig:fexp-moti2} shows the SM and DRAM bandwidth utilization of the decode phase for LLaMA3-8B under different configurations. As shown, across various configurations of batch size ($bs$) and sequence length ($seqlen$), the decode phase exhibits an average DRAM bandwidth utilization of $85\%$, while the average SM utilization remains as low as 40\%. Moreover, as $seqlen$ increases, both DRAM and SM utilization remain largely unchanged. 


To further analyze this imbalance, we dive into the characteristic of major kernels in decode phase. 
We focus on analyzing the matrix multiplication, which are the major kernels in LLM.
In modern GPUs, these matrix multiplications are accelerated using Tensor Cores, whose minimal processing unit is a $16 \times 16$ tile (for FP16 on A100 and Ada6000 GPUs).
As a result, each output matrix is partitioned into tiles of this size, with each tile computed by a single warp (32 threads, the minimum execution unit on SM).
On this basis, we estimate the total number of warps per matrix multiplication kernel as $Num_{warp} = (M / 16) \times (N / 16)$, where $M$ and $N$ are the row and column dimensions of the output matrix.



In the decode phase of LLMs, $M$ typically corresponds to the batch size $B$, and $N$ corresponds to the hidden dimension.
For LLaMA3-8B, where $H = 4096$, and assuming $B \leq 64$, the maximum number of warps is $64 \times 16 = 1024$.
Given that the Ada6000 has 142 SMs and each SM can accommodate up to 32 warps (ignoring shared memory and register constraints), the total warp capacity is $142 \times 32 = 4544$.
This analysis indicates that when $B < 64$, the decode phase cannot fully saturate the GPU’s SM resources, which explains the observed low SM utilization.



\subsubsection{The Opportunity}

\begin{figure}
\centering
\includegraphics[width=\linewidth]{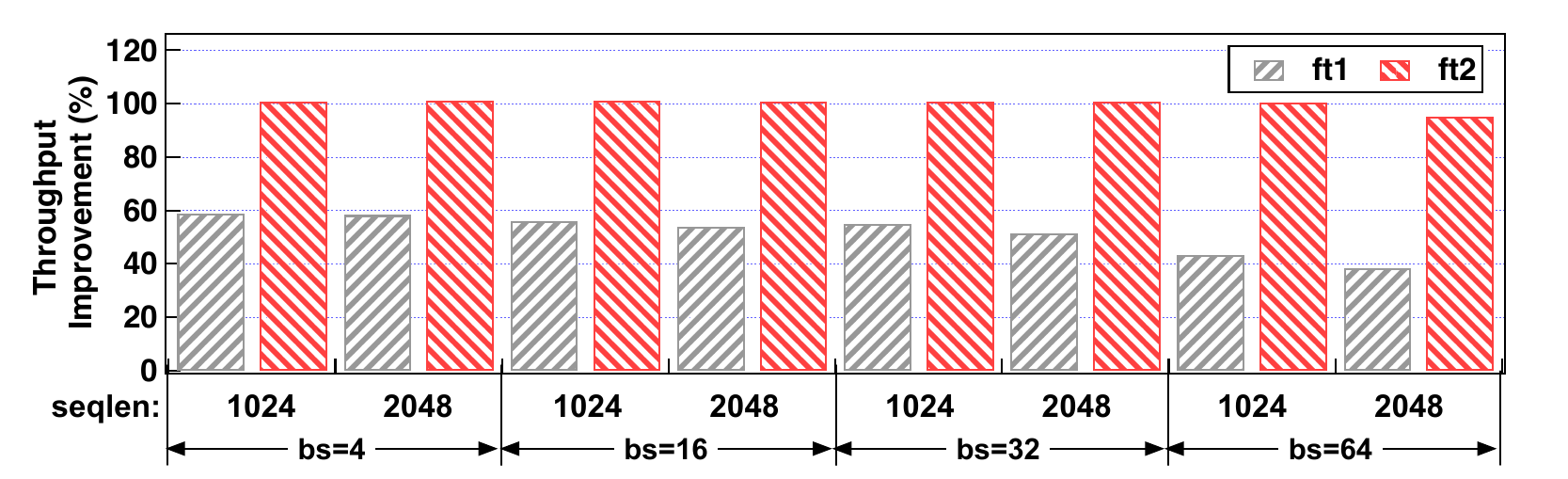}
\caption{The throughput improvement of finetune tasks under different configurations.}
\label{fig:fexp-moti3}
\end{figure}

A natural approach to addressing low SM utilization under runtime load fluctuation is to allocate a portion of SMs to a best-effort task with high compute intensity. In this context, we identify PEFT-based (parameter-efficient finetuning) LLM finetune tasks as promising candidates for co-location with LLM inference services.

To evaluate the potential throughput benefit, we construct two simplified finetune tasks, $ft_1$ and $ft_2$, each based on a single transformer layer from LLaMA3-8B. $ft_1$ performs only forward computation, while $ft_2$ performs only backward computation. In the baseline setup, the inference and finetune tasks are executed on two separate GPUs, with each task occupying an entire GPU. In the co-located setup, we partition GPU SMs between inference and finetune tasks across both two GPUs using spatial sharing. We set a QoS target of 40ms for TPOT of the decode phase.

\autoref{fig:fexp-moti3} reports the throughput improvement brought of $ft_1$ and $ft_2$ across various configurations (different batch sizes and sequence lengths) brought by co-location. In each setting, we manually tune the SM ratio between the two tasks to ensure that the inference QoS target is met while maximizing finetune throughput. As shown, co-location yields up to a $101.2\%$ improvement in finetune throughput without compromising inference QoS.

These results reveal a key opportunity: {\bf LLM finetune task is an effective co-location candidate for mitigating low GPU utilization during runtime load fluctuations in inference workloads.}


\subsection{Challenges}

Despite the potential to improve system throughput and resource utilization, co-locating LLM inference and finetune tasks presents significant challenges due to severe memory capacity and bandwidth contention. A single GPU often cannot accommodate the memory demands of both tasks, and under spatial sharing, they compete for DRAM bandwidth, leading to interference and performance degradation.

To address memory capacity contention, memory allocation for inference must be prioritized due to its strict QoS constraints. Consequently, the finetune task can only utilize memory not currently used by inference. Simultaneously, to address memory bandwidth contention, it is critical to accurately predict performance under varying resource partitioning to ensure QoS during co-execution.

Although many prior works\cite{rhu2016vdnnvirtualizeddeepneural,10.1145/3373376.3378530,10.1145/3200691.3178491,10.1145/3373376.3378505,zero} have explored memory management for training task under limited capacity, these solutions typically focus on {\bf intra-task memory management}, while the co-location scenario requires the {\bf inter-task memory management.} Specifically, prior works assume the exclusive and static memory usage, and adopt offline method to search the memory management strategy. However, LLM serving systems generally pre-allocate all GPU memory to avoid allocation overhead, leaving no statically available memory for other tasks. Moreover, inference memory usage is highly dynamic, making the memory available to co-located tasks both unpredictable and transient. Therefore, it is non-trivial to enable the LLM finetune task to utilize the unused memory of LLM inference service.

In addition, prior works\cite{gputlet,usher,mudi, orion} on scheduling under spatial co-location generally fall into two categories: those that assume contention is negligible or tolerable, and those that rely on offline profiling to identify low-interference task pairs. However, both are insufficient for co-locating LLM inference and finetune tasks. First, inference workloads are highly sensitive to interference; our measurements show that co-execution with finetuning can cause up to 40\% degradation, risking QoS violations. Second, inference performance is highly dynamic—varying with request arrival patterns, sequence lengths, and batch sizes—making static co-location strategies ineffective. These challenges make it particularly difficult to predict performance degradation under multi-factor contention, which is essential for effective co-location.

\section{Overview}

\subsection{Insights}

To address the above two challenges, we derive two key insights:
\begin{itemize}
[leftmargin=*]
    \item The reserved KV cache space allocated for LLM inference can be repurposed to accommodate the memory demands of LLM finetune tasks. This enables finetune tasks to opportunistically utilize unused GPU memory without interfering with inference memory requirements.
    \item The resource usage of LLM finetune task is relatively stable due to its fixed batch size, allowing its performance interference on inference to be quantified via linear regression based on its SM usage. This enables accurate performance prediction under co-execution.
\end{itemize}

\subsection{\sysname{} Design}

\begin{figure}
\centering
\includegraphics[width=\linewidth]{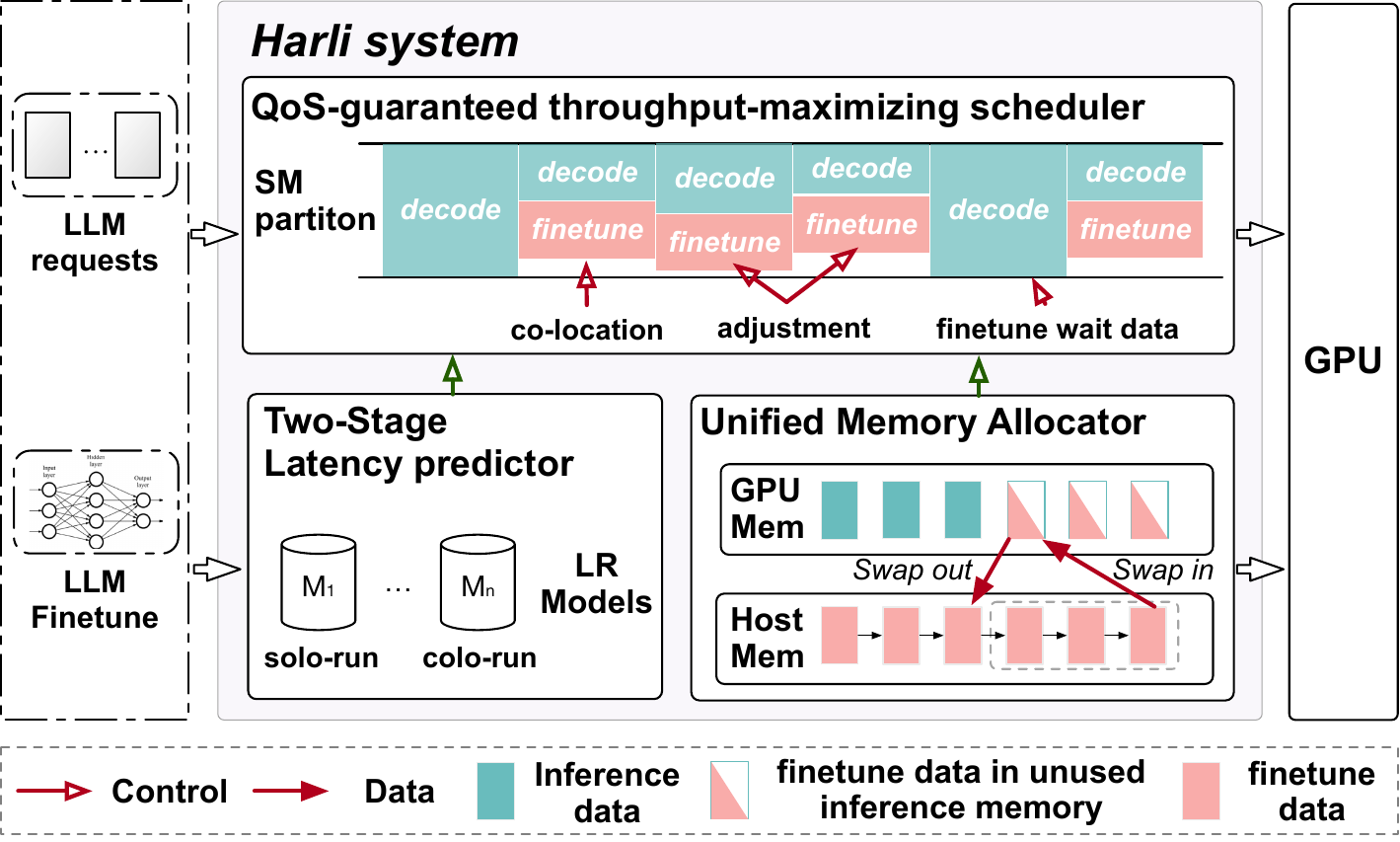}
\caption{System overview.}
\label{fig:overview}
\end{figure}

Building on these two insights, we propose \sysname{}, an LLM serving system that efficiently co-locates inference and finetune tasks on the same GPU. \autoref{fig:overview} illustrates the design of \sysname{}, which consists of three main components: a {\it unified memory allocator}, a {\it two-stage latency predictor}, and a {\it QoS-guaranteed throughput-maximizing scheduler}. 


At runtime, memory allocations from both tasks are managed by the unified memory allocator. This allocator repurposes unused KV cache memory from the inference service—via virtual memory mapping—for tensor allocations of the finetune task. Since the available memory is often insufficient to hold all tensors at once, the allocator supports window-based memory swapping. As shown in the figure, the finetune task determines the window size based on currently available memory. After completing the computation for one transformer layer, the allocator evicts the layer’s weights to host memory and loads the next layer into GPU memory. If the inference service requires memory that are occupied by the finetune task, the allocator dynamically shrinks the window and reclaims memory for inference.

Meanwhile, the latency predictor builds latency models for both solo-run and co-located scenarios. Specifically, the predictor first predicts the decode latency of the inference task under the current setup (batch size and sequence length) across different SM ratios. It then predicts the latency degradation under co-location with a finetune task. Since the batch size of the finetune task is static, its performance impact is relatively stable. The decode latency under co-location could be accurately predicted using a linear regression model based on SM ratio of LLM finetune and the solo-run performance of LLM inference.

Using these predictions, the online scheduler selects an SM partitioning that satisfies the QoS constraints for inference while maximizing the throughput of the finetune task. SM partitioning is implemented using CUDA’s GreenContext. At the start of each decode execution, the latency predictor estimates the expected latency for the next token. If a QoS violation is likely, the scheduler reconfigures the SM partitioning after the current token completes. 


Note that, the core contributions of \sysname{} are two lightweight yet effective mechanisms for inter-task memory sharing and performance prediction under spatial co-location. These techniques enable underutilized memory and compute resources in LLM serving systems to be reused by co-located tasks. While we focus on finetune tasks in this work, the approach generalizes to any PyTorch-based workload. We plan to open-source \sysname{} after publication.
\section{Unified Memory Allocator}


\subsection{Memory Usage Analysis and Principles}

To enable LLM finetune tasks to utilize memory unused by LLM inference, we first analyze the memory usage characteristics of the LLM inference. 

In LLM serving systems, memory usage during inference typically falls into three categories: model weights, intermediate activations, and KV cache. Model weights and activations are allocated through the default PyTorch allocator, while KV cache memory is managed by the serving system itself. Specifically, the serving system pre-allocates a large memory region and assigns KV cache slots to tokens using an index-based allocation strategy. This design significantly reduces runtime memory allocation overhead.

In contrast, PEFT-based LLM finetune tasks also utilize memory for three types of data: weights, activations, and gradients—all of which are allocated via the default PyTorch allocator. Notably, PEFT only updates a subset of model parameters, meaning that weights and activations can be divided into two categories: those that require gradients and those that do not. 


This analysis reveals a substantial mismatch between the KV cache management strategy used by the serving framework and PyTorch’s default memory allocation mechanism. Even if both inference and finetune tasks share the same allocator instance, the finetune task cannot access memory that has been reserved—but not yet used—for KV cache. As a result, we propose a unified memory allocator to support fine-grained inter-task memory management. This unified allocator must satisfy three key design principles.
\begin{itemize}
[leftmargin=*]
    \item The allocator must be efficient, preserving the fast KV cache allocation required for latency-sensitive inference.
    \item The allocator must be general-purpose, supporting shared memory management between LLM inference and general PyTorch-based tasks.
    \item The allocator must enable dynamic inter-task coordination, as inference memory usage fluctuates at runtime. 
\end{itemize}

\subsection{Unified Memory Management}



\begin{figure}
\centering
\includegraphics[width=\linewidth]{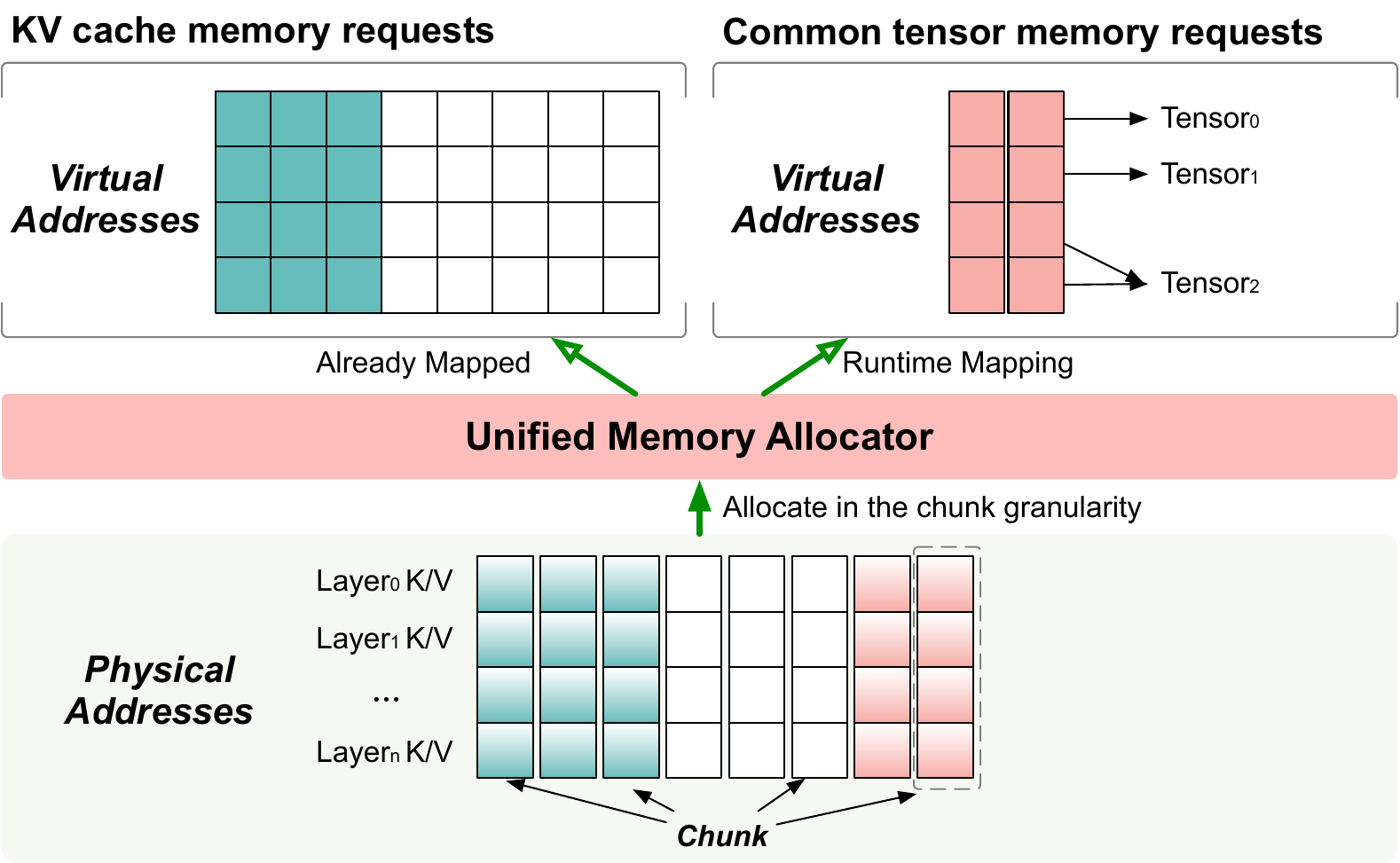}
\caption{The unified memory allocator.}
\label{fig:allocator}
\end{figure}

To enable inter-task memory management, \sysname{} extends the LLM serving framework to simultaneously support both the LLM inference service and the LLM finetune task. On this foundation, the unified memory allocator replaces the default PyTorch allocator, allowing any general-purpose tensor to reuse KV cache memory that has been reserved by the inference service but not yet utilized. \autoref{fig:allocator} illustrates the design of the unified memory allocator.


Specifically, the allocator leverages CUDA’s virtual memory management APIs, which allow a physical memory region to be mapped to multiple virtual addresses. Using these APIs, the unified memory allocator first pre-allocates all GPU memory into a memory pool. As shown in \autoref{fig:allocator}, this pool is initially organized into memory blocks that match the KV cache access pattern. The physical addresses of these blocks are mapped to the virtual addresses used by the inference service’s KV cache, preserving the zero-overhead allocation behavior ({\bf Principle 1}). 

Meanwhile, the allocator supports mapping unused blocks to new virtual addresses and serves them to other memory requests, such as those from the finetune task or activations and weights from the inference service. In this way, general-purpose tensors can safely and efficiently reuse memory originally reserved for KV cache without interfering with inference performance ({\bf Principle 2}).




The memory pool is organized as a 2D grid of memory blocks to align with the KV cache access pattern. Each block represents the KV cache of a single token in a single transformer layer, and each column corresponds to the KV cache of a specific token across all layers. However, due to limitations of CUDA’s virtual memory management APIs, the minimum allocatable block size is 2MB, while each KV cache entry requires a minimal amount of memory (e.g., 2KB per token per layer for Llama3-8B)
To address this mismatch and improve memory efficiency, we implement a two-level memory management strategy.



As shown in \autoref{fig:allocator}, the memory pool is divided into chunks, each with $layerNum \times 2$ blocks, where $layerNum$ is the transformer layers of the LLM inference. When the KV cache requests memory, a chunk is allocated to serve the KV cache entries for thousands of tokens. For general-purpose tensors, the allocator also assigns memory at the chunk level. While one tensor generally could not fully utilize the chunk, and the subsequent tesnor allocations could also be served by the chunk. Each tensor occupies an integer number of memory blocks. Once all blocks within a chunk are released, the chunk is returned to the memory pool for reuse.


\subsection{Memory Management Strategies}

Although the unified memory allocator enables inter-task memory management, the GPU memory is still insufficient to hold all tensors required by the LLM finetune task. To address this limitation, we design a window-based memory management strategy that performs tensor swapping between GPU and host memory, enabling the finetune task to execute correctly under constrained memory conditions.


However, not all tensors can be safely swapped from GPU to CPU memory due to PyTorch’s dynamic computation graph design. During the forward pass, PyTorch incrementally builds a computation graph that is later used to execute the backward pass. Tensors that require gradient updates must remain in GPU memory during the forward pass, as swapping them out would break the computation graph and result in incorrect or failed gradient computation.


In PEFT-based LLM finetune tasks, tensors can be broadly categorized into weights and activations, each further classified based on whether they require gradient updates. For activations, PyTorch already includes optimization strategies to discard intermediate activations that are not needed for backward computation. For weights, we distinguish between frozen weights (which remain unchanged) and trainable weights (which are updated during training). Based on this analysis, we retain all activations and target weights in GPU memory, and only manage frozen weights via swap-in and swap-out.

Specifically, the window-based memory management operates as follows. During inference execution, the allocator determines the window size based on the unused memory chunks. As the LLM finetune task completes the computation of one transformer layer, the frozen weights of that layer are evicted to host memory, and the frozen weights of the first layer outside the window are prefetched into GPU memory. Simultaneously, the finetune task proceeds to the computation of the next layer. To improve efficiency, we employ two CUDA streams to overlap computation and memory transfer.



\subsection{Inter-task Memory Coordination}


When the inference service demands additional memory, the finetune task must relinquish the corresponding portion. However, this memory cannot be released immediately. In particular, the release latency is the time required to swap out the frozen weights of a transformer layer to host memory.

To ensure that inference can obtain memory immediately when needed, we always pre-reserve a portion of KV cache memory during the inference execution ({\bf Principle 3}). This reserved memory can be conservatively estimated as: $Memory_{reserved} = (T / 50) * max_{bs} * Mem_{kv}$. Here, $T$ is the time to swap out one transformer layer in the finetune task, 50 ms is the decode phase QoS target, $\max_{bs}$ is the maximum supported batch size for inference, and $Mem_{kv}$ is the per-token KV cache memory requirement. This reserved threshold is typically smaller than the memory footprint of a single layer’s frozen weights, making it feasible.


\subsection{Memory Optimizations}

Based on the above design, \sysname{} enables efficient inter-task memory management. However, we observe that both tasks generate a large number of small tensor allocations.
For example, in a finetune configuration with batch size 2 and sequence length 1024, we observe over 5k tensors allocations smaller than 2MB during one training iteration.
Serving these small tensors using the default 2MB block granularity incurs significant allocation overhead and leads to severe memory fragmentation.

To address this, we find that most of these small tensors correspond to intermediate activations, whose allocation and release follow predictable temporal patterns. Then, we design a separate small-tensor memory pool. In this pool, all physical memory is directly mapped to virtual addresses, and the allocation granularity is reduced to 2KB. We implement a buddy allocation scheme within this pool to efficiently manage dynamic allocation and release of small tensors. At system initialization, we profile the activation memory demand of both the inference and finetune tasks and statically set the size of this small-tensor pool accordingly.

\section{Two-stage Latency Predictor}
\label{sec:Predictor}

In this section, we present the two-stage latency predictor designed to estimate decode phase latency when co-locating the LLM decode instances with a finetune task. Specifically, we first predict the solo-run decode latency, and then predict the decode latency at co-location. 



\subsection{Prediction for Solo-run Latency}


Although many existing efforts optimize LLM serving systems, they typically rely on extensive profiling to characterize the runtime behavior of inference workloads. To develop a lightweight yet accurate performance model, we begin by measuring the decode latency of LLM inference service.

\begin{figure}
\centering
\includegraphics[width=\linewidth]{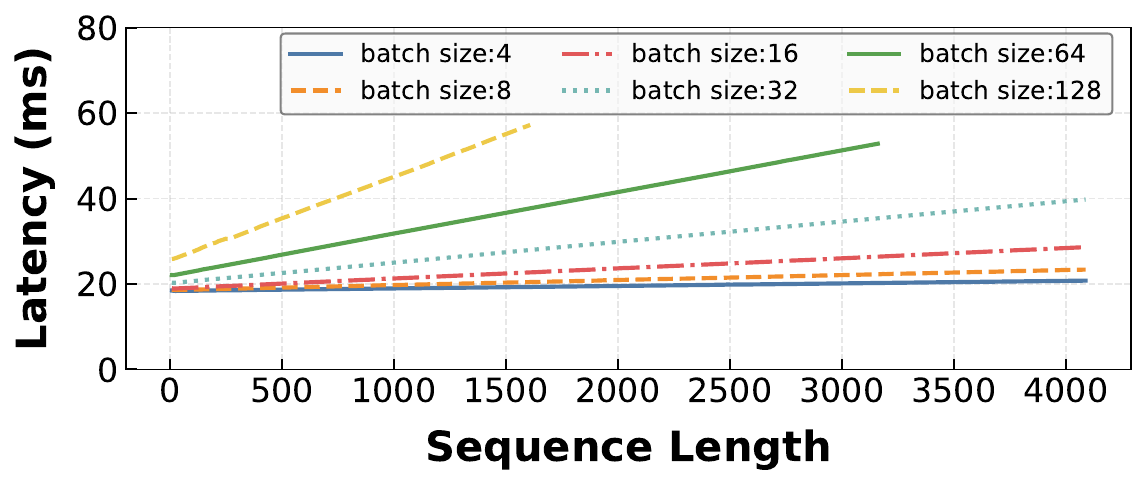}
\caption{Decode latency under different bs and seqlen.}
\label{fig:5-1-figure1}
\end{figure}

\autoref{fig:5-1-figure1} shows the decode latency of Llama3-8B under different batch sizes and sequence lengths with full SM usage. From this figure, we make three key observations. First, decode latency increases linearly with sequence length across all batch sizes due to the increasing workloads. Second, for batch sizes less than or equal to 4, the latency curves are nearly identical. This effect arises from padding strategies employed by modern serving frameworks to align workloads for Tensor Core acceleration. Third, when batch size exceeds 4, the latency curves exhibit similar slopes when the input changes to batch size multiply sequence length. 


Based on these observations, we propose a LR-based (Linear Regression) latency model for decode latency at solo-run execution. The model predicts decode latency as:

\begin{equation}
\small
\begin{aligned}
    Latency_{Decode} = \text{bs} \cdot b_0 + c_0 + \text{bs} \cdot k_0 \cdot \text{seqlen}
\end{aligned}
\label{eq:decode_predict}
\end{equation}

where $bs$ is the current batch size, $seqlen$ is the output length, and $b_0$, $c_0$, $k_0$ are coefficients determined using profiling. Experimental results show that this model achieves a mean prediction error of just 3\%, proving its effectiveness for fast and practical runtime latency prediction.


\begin{figure}
\centering
\includegraphics[width=\linewidth]{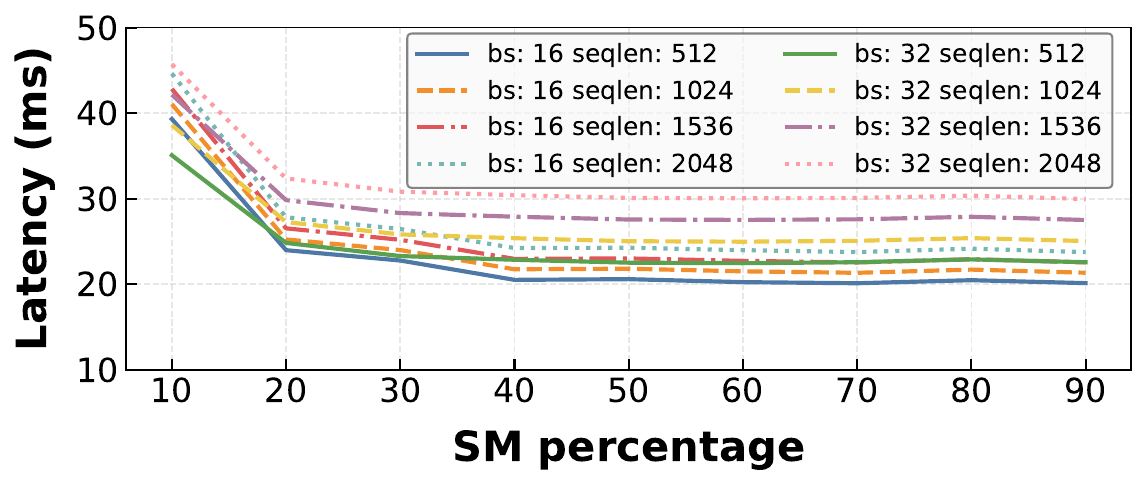}
\caption{Solo decode latency under different SM count.}
\label{fig:5-1-figure2}
\end{figure}

Further, we profile the decode latency of LLM inference under varying SM allocation ratios. Specifically, we randomly select a range of ($bs$, $seqlen$) configurations and measure their decode latency while allocating between 10\% and 100\% of the GPU SMs. As shown in \autoref{fig:5-1-figure2}, all configurations exhibit sublinear performance scaling with increasing SM ratio. These performance lines show great differences. Therefore, it is hard to construct a precise performance model for all batch sizes and sequence lengths.

Since SM ratios are typically discretized in coarse-grained steps (e.g., 5\% or 10\%), we construct the performance model for each SM resource configuration using \autoref{eq:decode_predict}. Note that constructing the latency model for solo-run cases does not require exhaustive profiling. Specifically, profiling with just three representative batch sizes is sufficient for accurate modeling (evaluated in \autoref{sec:acc} and \autoref{sec:overhead}).

\subsection{Prediction for Co-run Cases}

\subsubsection{Constructing the Model}
Assuming a minimum granularity of 10\% SM partitioning, \sysname{} must consider up to 45 possible allocation combinations between the two tasks. Identifying the optimal one that satisfies the QoS target of the inference service while maximizing the throughput of the finetune task is a great challenge.

\begin{figure}
\centering
\includegraphics[width=\linewidth]{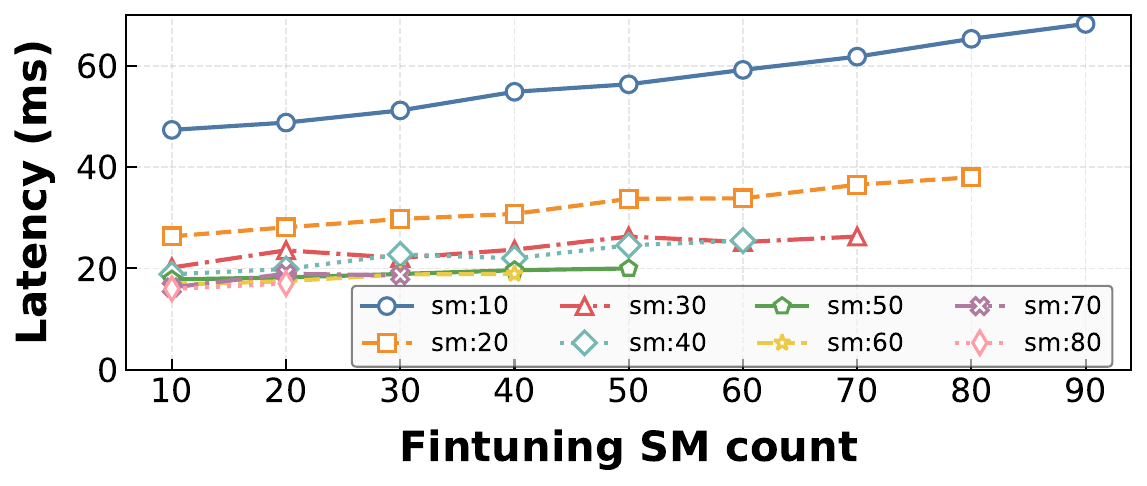}
\caption{Colo decode latency under different SM allocation combinations.}
\label{fig:prediction step2}
\end{figure}

To address this, we first collect the decode latency under different SM allocation combinations. As shown in \autoref{fig:prediction step2}, each lines represent the decode latency of LLM inference using the same SM ratio with LLM finetune task using increasing SM ratios. As shown, all the lines have similar slopes. These empirical observations motivate the construction of an contention-aware latency model using \autoref{eq:predict_spatial}. This models predicts the decode latency $Latency_{colo}$ under different SM allocations as:
\begin{equation}
\small
\begin{aligned}
    Latency_{colo} = (SM_{infer} \cdot b_1 + SM_{ft} \cdot k_1) * Latency_{Decode-sm}
\end{aligned}
\label{eq:predict_spatial}
\end{equation}

where $SM_{infer}$ is the SM ratio used by LLM inference, $SM_{ft}$ is the SM ratio used by LLM finetune, $Latency_{Decode-sm}$ is the decode latency under the $SM_{infer}$ allocation and $b_1$, $k_1$ are coefficients determined using profiling. This model achieves a prediction accuracy of 95\%, confirming its effectiveness across different resource configurations. Note that experimental results show a single model can effectively capture both forward and backward contention, owing to the similarity in their underlying computation operators.

\subsubsection{Theoretical Analysis}

In addition to empirical modeling, we further prove the model construct in a theoretical way. Specifically, we abstract the interference between LLM inference and finetune workloads as a two-task memory bandwidth contention model, where LLM inference maintains a fixed memory access rate, and LLM finetune exhibits varying levels of memory traffic. The goal is to predict the performance of LLM inference as a function of LLM finetune's increasing memory pressure. 

Under this setting, we model the shared memory bandwidth with total capacity $B$ (measured in memory accesses per second). LLM inference service issues memory accesses at a fixed rate $f_{infer}$, while LLM finetune task exhibits a variable memory access rate $f_{ft}$. When executed individually, both tasks satisfy $f_{infer} \leq B$ and $f_{ft} \leq B$. However, when co-located, their combined demand $f_{infer} + f_{ft}$ may exceed the available bandwidth, i.e., $f_{infer} + f_{ft} > B$.

In the case of contention, the shared bandwidth is distributed proportionally to each task’s demand. Accordingly, the effective memory processing rate for LLM inference, denoted $r_{infer}$, is given by:
\begin{equation}
\small
\begin{aligned}
    r_{infer} = \frac{B \times f_{infer}}{f_{infer} + f_{ft}}
\end{aligned}
\label{eq:infer-access}
\end{equation}


To quantify the latency degradation experienced by LLM inference, we note that execution latency is inversely proportional to its effective processing rate. The original latency under solo execution is proportional to $1 / f_{infer}$, while the latency under contention is proportional to $1 / r_{infer}$. Therefore, we could get the latency slowdown $Factor_{slowdown}$ and the decode latency under the co-location $Latency_{colo}$.
\begin{equation}
\small
\begin{aligned}
    Factor_{slowdown} &= \frac{1 / r_{infer}}{1 / f_{infer}} = \frac{f_{infer}}{r_{infer}} = \frac{f_{infer} + f_{ft}}{B} \\
    Latency_{colo} &= Factor_{slowdown} * Latency_{Decode-sm} \\
    &= \frac{f_{infer} + f_{ft}}{B} * Latency_{Decode-sm}
\end{aligned}
\label{eq:infer-degrad}
\end{equation}

This result shows that the decode latency grows linearly with the total memory demand, which is the increasing SM allocations of LLM finetune task.

\subsubsection{Perceive the Finetune Performance}

While we can accurately predict the performance of decode latency under co-location, it is significantly more difficult to model the impact of inference configurations on the throughput of finetuning. This is because the finetune task could operate with a relative fixed configuration (static batch size), whereas inference workloads exhibit highly dynamic behavior. 


Despite this modeling challenge, we make a critical empirical observation: the throughput of the finetune task tends to be highest when the inference task is operating near—but not exceed—its QoS threshold. This is because, when inference latency closely approaches its QoS target, the inference workload is not over-consuming memory bandwidth, which leaves more bandwidth available for the finetune task. Based on this observation, our scheduler simply selects the SM configuration for inference that yields the closest latency to the QoS target. This implicitly maximizes the remaining bandwidth for finetuning and thus improves its throughput.


Furthermore, we find that it is often unnecessary for inference and finetune tasks to jointly utilize all available SMs. Once memory bandwidth becomes the bottleneck, increasing compute allocation does not improve throughput. 
\section{QoS-guaranteed Throughput-maximizing Scheduler}

In this section, we describe the mechanism used to schedule LLM inference services and finetune tasks.

\subsection{Determining the Scheduling Unit}

At runtime, the scheduler must coordinate SM usage between the inference and finetune tasks to ensure inference QoS while maximizing the GPU utilization. However, the two tasks operate with different scheduling units, making the coordination difficult.

Specifically, the scheduling unit for LLM inference is per-token generation during the decode phase, which operates under a strict QoS target, such as 50 ms. In contrast, the default scheduling unit for the finetune task encompasses an entire forward and backward pass, which is significantly longer in duration and cannot be preempted once launched.

Faced with this mismatch, the LLM finetune task must be divided into finer-grained scheduling units. A natural idea is to treat each transformer layer as a separate scheduling unit. However, PyTorch-based training workflows do not expose such granularity at the framework level. In typical usage, the backward pass is submitted via a single call (e.g., loss.backward()), which triggers all the gradient computation in C++ backend—making it infeasible to interleave or suspend at intermediate layers from the Python side.


To overcome this limitation, we partition the LLM model into multiple submodels, each corresponding to a single transformer layer. During the backward pass, each submodel’s output tensor is passed explicitly as the input to the next submodel, enabling explicit, layer-wise gradient computation and scheduling at the framework (PyTorch) level. This approach enables the finetune task to be scheduled in a layer-wise manner, thus supporting fine-grained, cooperative scheduling with the inference task.

Further, we reduce the batch size of the finetune task by splitting the mini-batch into micro-batches, ensuring that the execution time of each forward or backward submodel remains around 10 ms. This guarantees that the finetune scheduling unit is shorter than the inference scheduling window, enabling responsive and low-overhead synchronization between the two tasks.

\subsection{Runtime Scheduling}

At the beginning of each decode phase, the scheduler first estimates the solo-run latency of LLM inference under the current batch size and sequence length. It then evaluates the co-located latency of inference when sharing GPU resources with the finetune task under various SM allocation ratios. Based on these latency predictions, the scheduler selects the SM allocation plan that satisfies the QoS constraint for inference while maximizing overall GPU utilization.

When a new request arrives or the system proceeds to the next decode phase, the scheduler re-evaluates whether the current SM allocation might cause QoS violations. If a violation is predicted, the scheduler pauses the finetune task and recomputes a new SM allocation plan that ensures QoS for inference. If the current allocation is still valid, both inference and finetune tasks continue executing.

In parallel, due to the fact that the data transfer time for one transformer layer in the finetune task exceeds its compute time, the finetune task may temporarily become computation-starved. When this occurs, and the finetune task has no ready work to execute, the scheduler temporarily reclaims all SMs for inference during the next decode phase. Once the finetune task resumes computation, the scheduler redistributes SMs between the two tasks.

Through this adaptive spatiotemporal scheduling strategy, the system ensures high GPU utilization at all times while maintaining inference QoS guarantees.
\section{Implementation}
We implement the prototype of \sysname{} based on SGLang\cite{NEURIPS2024_724be447}, a state-of-the-art LLM serving system, and LlamaFactory\cite{zheng2024llamafactory}, a state-of-the-art LLM finetuning framework, with 4,000 lines of code: 2,500 lines of C++ for the unified memory allocator, 200 lines of C++ for integrating GreenContext\cite{greencontext} into PyTorch\cite{paszke_pytorch_2019} to enable spatial co-location, and 1,300 lines of Python for the latency predictor and task scheduler.
GreenContext is a low-level compute resource management mechanism that provides intra-process, fine-grained compute partitioning across different CUDA streams~\cite{Nvidia2025cudastream}.

Since both SGLang and LlamaFactory are PyTorch-based systems, \sysname{} initializes separate model instances within a single process and allocates distinct SM partitions to them via GreenContext.
As a result, \sysname{} automatically supports the co-location of inference and PEFT tasks for any mainstream LLMs supported by SGLang and LlamaFactory.

\begin{figure*}
\centering
\includegraphics[width=\linewidth]{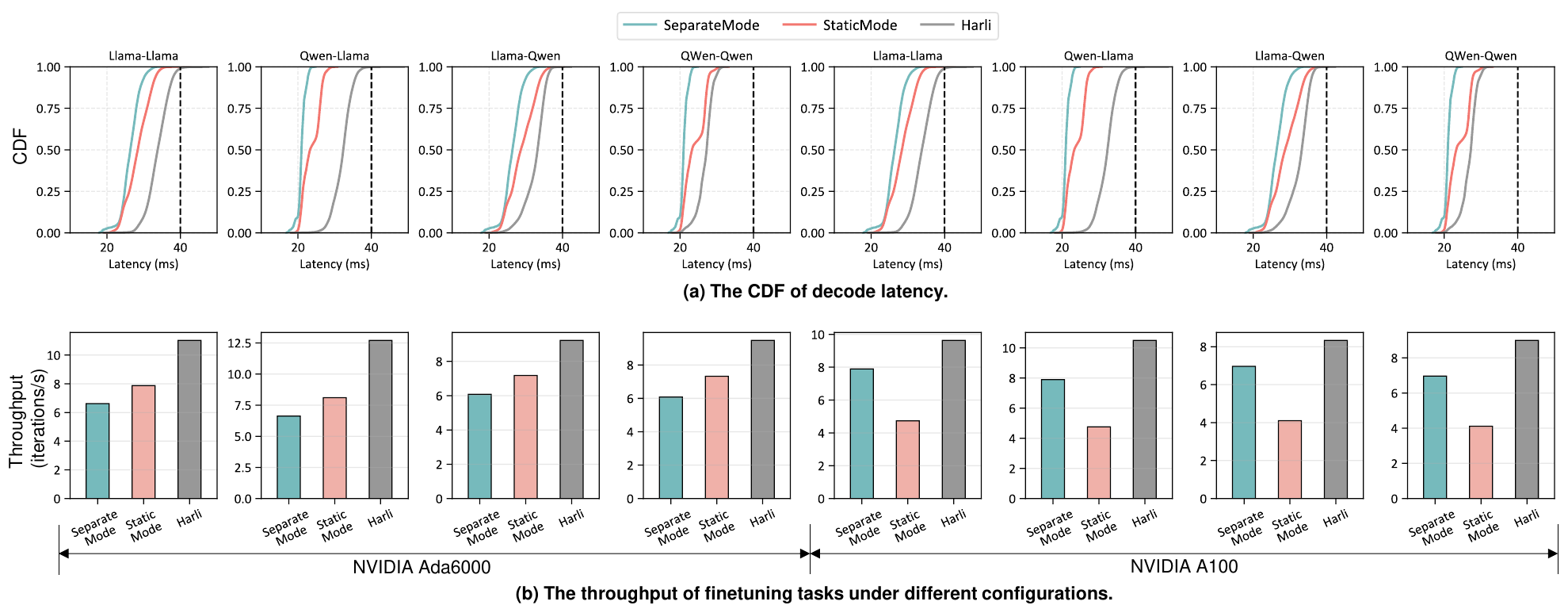}
\caption{Comparison between \sysname{} and two baselines, SeparateMode and StaticMode, in improving throughput of finetuning tasks while maintaining QoS for inference requests. The caption of each subfigure, X-Y, indicates using model X for inference and model Y for finetuning.}
\label{fig:throughput improvement}
\end{figure*}

\section{Evaluation}
In this section, we evaluate \sysname{} in improving throughput of finetune tasks while promising the QoS requirements of inference tasks.
\subsection{Experimental Setup}
\label{sec:lab8_1}

\begin{table}
\footnotesize
\centering
\caption{\label{tab:system} Hardware and software specifications.}
\begin{tabular}{l|l}
\hline
\hline
 & Configuration
\\ \hline
CPU              & Intel Xeon Gold 6530 64-core \\ \hline
GPU              & Nvidia Ada6000 48GB \& Nvidia A100 40GB\\ \hline
Nvidia Driver    & 570.124.06                     \\ \hline
Runtimes         & CUDA 12.8         \\ \hline
AI framework     & Pytorch 2.5.1 \& SGLang 0.4.0           \\ \hline
\hline
\end{tabular}
\end{table}


\paragraph{Testbeds.} \autoref{tab:system} shows the hardware configuration used in our experiments. All evaluations are conducted on Nvidia Ada6000 and Nvidia A100 GPUs. Notably, \sysname{} does not rely on any hardware-specific features of Ada6000 or A100 and can be deployed on other GPUs without modification.

\paragraph{Workloads.} We evaluate \sysname{} using two widely adopted LLMs: LLaMA3-8B (abbreviated as Llama)\cite{meta-llama2023llama3} and Qwen2.5-7B (abbreviated as Qwen) \cite{qwen}, which serve as benchmarks for both LLM inference and finetuning. The inference service is deployed in PD disaggregation mode, built on top of the mainstream serving framework SGLang \cite{sglang}. At runtime, we allocate one GPU instance to the prefill phase and concentrate GPU utilization on the instance serving the decode phase. While the QoS target for the decode phase typically ranges from 40 ms to 80 ms, we adopt a more stringent target of 40 ms TPOT to evaluate \sysname{}. 
The workload is driven by a one-hour trace containing over 19000 requests
\cite{splitwise}.

\paragraph{Baselines.} We use the following two baseline approaches.
\begin{itemize}
[leftmargin=*]
    \item {\bf SeparateMode:} The LLM inference service and the LLM finetune task run on separate GPUs: one GPU is dedicated to the decode phase of inference, while the other is used exclusively for LLM finetuning.
    \item {\bf StaticMode:} The LLM inference and LLM finetune tasks are co-located on both two GPUs using spatial sharing. Specifically, on each GPU, both SMs and memory are statically partitioned: the inference task is allocated 60\% of the SMs and 60\% of the GPU memory, while the remaining resources are reserved for the finetune task.
\end{itemize}

\subsection{Improving Throughput}

In this subsection, we compare \sysname{} against two baseline approaches. We use iterations per second multiplied by batch size as the throughput metric. 
We focus on reporting finetune throughput improvements, while maintaining QoS for all inference requests.


\autoref{fig:throughput improvement}(b) compares the throughput of LLM finetune tasks across three setups: \sysname{}, SeparateMode, and StaticMode. As shown in the figure, on Nvidia Ada6000, \sysname{} achieves an average throughput improvement of 46.2\% (up to 92.0\%) over SeparateMode, and an average of 75.1\% (up to 120.5\%) over StaticMode. \sysname{} consistently improves throughput across all co-location pairs and configurations by fully utilizing underutilized SM resources and memory capacity. In contrast, SeparateMode fails to leverage these idle resources, and StaticMode can only partially exploit them due to its fixed resource partitioning. 

We observe that \sysname{} achieves higher throughput on the Nvidia Ada6000 GPU compared to the Nvidia A100. This improvement can be attributed to two key factors. First, the Ada6000 has more SMs—142 compared to 108 on the A100—providing greater opportunities for co-location and fine-grained SM partitioning. Second, the Ada6000 offers 48 GB of GPU memory, whereas the A100 provides only 40 GB. The additional memory reduces the frequency of memory swapping for the LLM finetune task, further improving efficiency. As a result, \sysname{} delivers higher overall throughput on the Ada6000 GPU.

Since GPU memory capacity is more constrained on A100, StaticMode statically allocate less memory to finetune tasks, leading to reduced throughput for finetuning tasks compared to SeparateMode on A100.


Note that \sysname{} configure the batch size of the LLM finetune task using micro-batching. Specifically, we set the batch size to 2 for both finetune models in StaticMode and \sysname{}. 
In contrast, we do not modify the batch size in the baseline systems, where the finetune batch size is set to 16. 

\subsection{Guaranteeing QoS}

We evaluate the decode latency of four co-location pairs on the Nvidia Ada6000 and A100. \autoref{fig:throughput improvement}(a)  presents the cumulative distribution function (CDF) distribution of runtime decode latency across these configurations. As shown in \autoref{fig:throughput improvement}(a), \sysname{} consistently meets the QoS target for the decode phase of LLM inference under all co-location scenarios. This is achieved by selecting the SM partitioning strategy based on the predicted decode latency under co-location, aiming to maximize throughput while satisfying QoS constraints. Furthermore, the decode latency under \sysname{} stays closer to the QoS target, which directly contributes to the increased throughput of the finetune task.

\subsection{Accuracy of The Duration Predictor}\label{sec:acc}

\begin{figure}
\centering
\includegraphics[width=\linewidth]{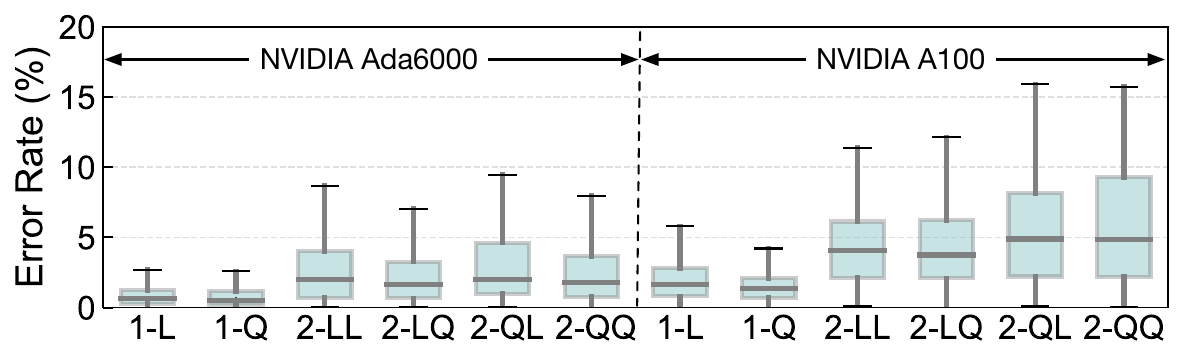}
\caption{Box plot of prediction error rates. Labels on the x-axis indicates the inference decoding latency prediction stage and models. Label 1-X denotes stage-1 (i.e. prediction for solo-run latency) using model X, and 2-XY indicates stage-2 (i.e. prediction for co-run cases) using model X for inference and model Y for finetuning. L indicates LLaMA models and Q denotes Qwen models.}
\label{fig:boxplot}
\end{figure}

In this subsection, we evaluate the latency prediction accuracy of the LLM inference service. 



In this experiment, we first evaluate the prediction accuracy of the linear regression (LR) models used for solo-run cases. We assess these models by applying them to \sysname{} running without the LLM finetune task. The bars labeled “1-X” in \autoref{fig:boxplot} show the prediction errors for the solo-run scenarios. The predicted decode latency deviates from the actual runtime by no more than 6\%, with an average error of less than 2\%.

We also evaluate the prediction accuracy of the linear regression (LR) models for the co-located cases. As shown by the bars labeled “2-XY” in \autoref{fig:boxplot}, the prediction error when co-locating inference and finetune tasks remains below 5\% on average across all tested configurations.

These results demonstrate that the two-stage latency predictor in \sysname{} is sufficiently accurate for estimating the decode latency of LLM inference under co-location, enabling informed and effective resource scheduling.

\subsection{Effect of Unified Memory Allocator}
\label{sec:lab8_5}

\begin{figure}
\centering
\includegraphics[width=0.95\linewidth]{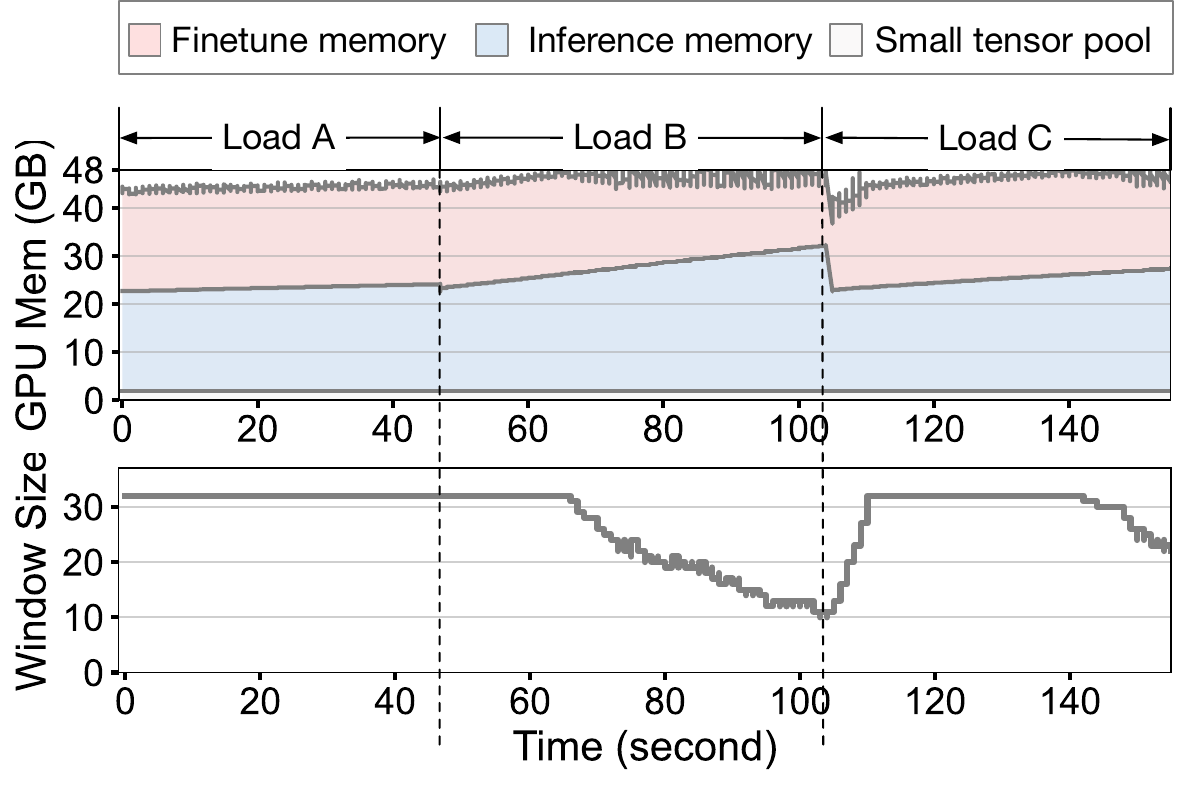}
\caption{The memory usage of both inference and finetune tasks, and the window size of the
finetune task. Loads a, b, c denote light load, heavy load, and medium load respectively.}
\label{fig:memory usage}
\end{figure}


In this subsection, we evaluate the effectiveness of the unified memory allocator by analyzing the memory usage distribution and the window size dynamics of the LLM finetune task during runtime. To do so, we employ a controlled load trace that simulates varying runtime demand: the system first experiences a light load (batch size = 8), followed by a heavy load (batch size = 42), and finally transitions to a medium load (batch size = 24) after the heavy load completes.

\autoref{fig:memory usage} illustrates the runtime memory usage of the Llama-Llama (i.e. using one Llama model for infernce, and using another Llama model for finetuning) co-location setup on an Nvidia Ada6000. The size of small-tensor pool remains constant, while the memory usage of the LLM inference service fluctuates in response to the dynamic request load. The LLM finetune task adapts accordingly, adjusting its memory footprint in response to these fluctuations. 

\autoref{fig:memory usage} also shows the window size adjustments of the LLM finetune task. As observed, increases in inference memory usage lead to corresponding reductions in the finetune window size, thereby lowering the finetune memory footprint.  This adaptive behavior aligns with the trends in the runtime memory usage. These results demonstrate that the unified memory allocator effectively coordinates memory usage between two tasks.


\subsection{Effect of Online Scheduler}
\begin{figure}
\centering
\includegraphics[width=0.95\linewidth]{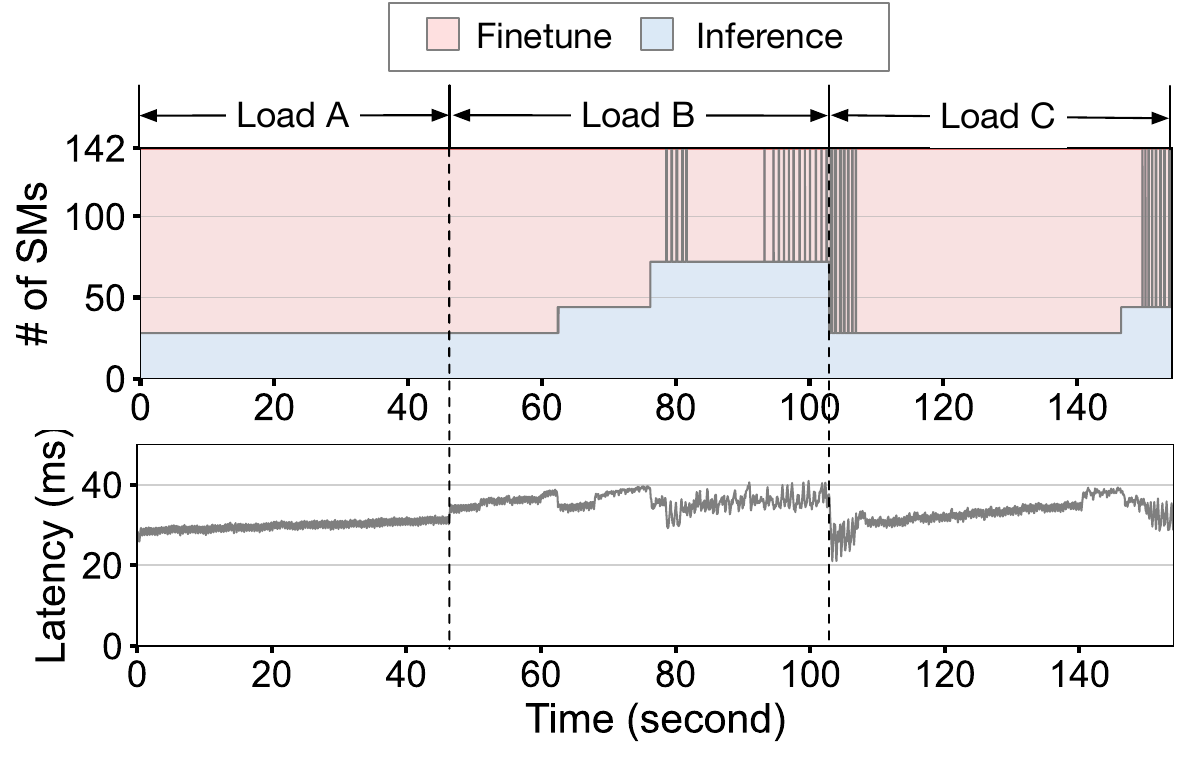}
\caption{The SM numbers allocated to inference and finetune tasks, and the decoding latency of the inference task.}
\label{fig:SM ratio}
\end{figure}


In this subsection, we evaluate the effectiveness of the QoS-guaranteed throughput-maximizing scheduler, using the same trace as \autoref{sec:lab8_5}. Since the LLM finetune task may stall while waiting for data swaps, the scheduler must allow the LLM inference task to preempt all SM resources to maintain high GPU utilization.

\autoref{fig:SM ratio} shows the runtime SM allocation between the inference and finetune tasks. As illustrated, when the finetune task is waiting for data, the inference task temporarily preempts all SM resources. Once the finetune task resumes execution with data ready, the scheduler dynamically redistributes SMs to allow shared execution. Resource adjustments occur approximately every 100 ms, and are brief—resulting in short-lived changes in the figure.

\autoref{fig:SM ratio} also presents the corresponding decode latency. As shown, the latency drops when the inference task is granted full SM access. This behavior confirms the effectiveness of \sysname{}’s online scheduler.

\subsection{Effect under Tensor Parallelism}

In this subsection, we evaluate the effectiveness of TP-based task co-location. Unlike \sysname{}, which requires maintaining multiple copies of model weights, \sysname{}-TP only needs a single copy of weights that are sharded across GPUs via tensor parallelism. Using the same experimental configuration as described in \autoref{sec:lab8_1}, we compare the performance of \sysname{}-TP against the original \sysname{}.

\sysname{}-TP achieves an average throughput improvement of 10.2\% over \sysname{} and 33\% over SeparateMode. This improvement is observed across all co-location pairs and configurations. The gains stem from reduced memory usage on each GPU for the inference task, which in turn lowers the memory swapping overhead for the finetune task. By reducing memory contention, \sysname{}-TP enables broader co-location opportunities and ultimately delivers higher system throughput.

\subsection{Overhead}
\label{sec:overhead}

The overhead of \sysname{} arises from three main sources:

First, \sysname{} requires offline preparation of latency prediction models. For the solo-run latency model, \sysname{} constructs 10 linear regression models, each corresponding to a different SM ratio. For each SM ratio, the model is trained using runs at batch sizes of 4, 16, and 64. In each batch size, the decode phase is executed once, covering sequence lengths up to 512 tokens. Overall, it takes approximately 6 minutes to construct solo-run models for both LLaMA and Qwen.

For the co-location latency model, \sysname{} builds a single model across all batch sizes and sequence lengths. It runs the decode phase at batch sizes of 4, 16, and 64, each with a maximum sequence length of 512, under 45 different SM allocation pairs. This process takes approximately 58 minutes. These profiling costs are acceptable, as both LLM inference and finetuning are long-running tasks where amortizing setup time is reasonable.

Second, \sysname{} introduces runtime latency prediction overhead. However, the LR-based prediction is highly efficient, taking only 5 us per invocation, which introduces negligible runtime overhead.

Third, \sysname{} allocates memory blocks at a fixed 2 MB granularity, which could potentially cause memory fragmentation. To address this, \sysname{} maintains a dedicated small-tensor memory pool for sub-2MB allocations. As a result, fragmentation remains minimal—typically under 100 MB in most runtime scenarios.

\section{Related Work}

\paragraph{Memory resource management.}
Prior work has extensively explored memory management for DNN and LLM training under constrained GPU memory. Techniques include data offloading, intermediate recomputation, and CPU-side execution. vDNN~\cite{rhu2016vdnnvirtualizeddeepneural} offloads intermediate activations to CPU DRAM, while SwapAdvisor~\cite{10.1145/3373376.3378530} uses genetic algorithms to optimize offloading decisions. SuperNeurons~\cite{10.1145/3200691.3178491} and Capuchin~\cite{10.1145/3373376.3378505} combine offloading and recomputation, dynamically selecting strategies based on tensor characteristics. These methods typically assume statically allocated GPU memory, enabling offline search for optimal policies. Meanwhile, ZeRO~\cite{zero} reduces memory usage by offloading optimizer states in distributed training.
\sysname{} adopts their host memory offloading mechanism to resolve inter-task memory contention, enabling safe harvesting of underutilized GPU resources in dynamic LLM serving scenarios.

\paragraph{Compute resource management.}
Many prior works have explored GPU resource sharing, broadly categorized into time-sharing and spatial-sharing. Time-sharing approaches like Baymax~\cite{baymax}, Gaina~\cite{gaina}, qCUDA~\cite{qcuda}, and vCUDA~\cite{vcuda} use API remoting to intercept CUDA calls and multiplex kernel execution across workloads. For spatial sharing, Nvidia provides hardware support via MPS~\cite{mps}, MIG~\cite{mig}, and GreenContext~\cite{greencontext}, enabling concurrent use of GPU SMs across applications. 
In this work, we adopt GreenContext for its precise SM partitioning, making it well-suited for co-locating LLM inference and finetuning tasks.

\paragraph{QoS guarantee under spatial multiplexing.}
Extensive researches have studied spatial sharing to co-locate tasks and improve throughput, focusing on two main aspects: (1) modeling interference and (2) selecting task pairs via offline profiling.
GPUlet~\cite{gputlet} uses linear regression on low-level metrics (e.g., L1 cache, bandwidth) to predict co-location interference. Usher~\cite{usher} profiles kernel-level compute/memory demands to co-locate complementary workloads. Mudi~\cite{mudi} scales this to cluster-wide scheduling using offline contention profiling, while Orion~\cite{orion} classifies operators by resource type to ensure stable co-location.
These techniques target traditional inference tasks and fall short for LLM workloads, where memory contention is significantly higher.
\section{Conclusion}

We propose \sysname{}, a system that enables efficient co-location of LLM inference and PEFT-based finetuning tasks to maximize GPU utilization. \sysname{} targets a key inefficiency in modern serving systems: the underutilization of decode instances due to their memory-bound behavior and dynamic workload variability. By leveraging the stable and lightweight nature of PEFT, \sysname{} safely co-locates finetuning with decode workloads without violating QoS guarantees. It introduces a unified memory allocator for inter-task memory sharing, a two-stage latency predictor for dynamic performance modeling, and a QoS-aware scheduler to balance resource allocation. Our evaluation shows that \sysname{} significantly improves overall throughput while maintaining strict inference latency targets. 


\bibliographystyle{ACM-Reference-Format}
\bibliography{sample-base}

\end{document}